# The mass of the young planet β Pictoris b through the astrometric motion of its host star

I.A.G. Snellen & A.G.A. Brown

*Leiden Observatory, Leiden University, Postbus 9513, 2300 RA Leiden, The Netherlands*

**The young massive Jupiters discovered with high-contrast imaging[1-4] provide a unique opportunity to study the formation and early evolution of gas giant planets. A key question is to what extent gravitational energy from accreted gas contributes to the internal energy of a newly formed planet. This has led to a range of formation scenarios from 'cold' to 'hot' start models[5-8]. For a planet of a given mass, these initial conditions govern its subsequent evolution in luminosity and radius. Except for upper limits from radial velocity studies[9-10], disk modelling[11], and dynamical instability arguments[12], no mass measurements of young planets are yet available to distinguish between these different models. Here we report on the detection of the astrometric motion of Beta Pictoris, the ~21 Myr-old host star of an archetypical directly-imaged gas giant planet, around the system's centre of mass. Subtracting the highly accurate Hipparcos[13,14]-Gaia[15,16] proper motion from the internal 3-yr Hipparcos astrometric data reveals the reflex motion of the star, giving a model-independent planet mass of $M=11\pm2$ $M_{Jup}$. This is consistent with scenarios in which the planet is formed in a high-entropy state as assumed by hot start models. The ongoing data collection by Gaia will in the near future lead to mass measurements of other young gas giants and form a great asset to further constrain early evolution scenarios.**

ESA's Hipparcos[13] (1989–1993) and Gaia[15] (2013–present) astrometric space observatories use the same measurement principles. Both contain two telescopes with lines of sight perpendicular to the spin axis of a continuously rotating spacecraft, which slowly precesses while maintaining a constant angle with respect to the spacecraft-Sun direction. The telescopes precisely time the crossing of stars in their field of view, providing accurate one-dimensional astrometric positions, which results in a rigid system of reference by combining the relative positions of the observed stars.

Hipparcos observed Beta Pictoris 111 times between 1990.0 and 1993.1, distributed over 35 spacecraft orbits. Each ~3-5 observations taken during one orbit have similar telescope scanning directions. While the initial data release delivered one position per orbit[13], with typical uncertainties in the range 1.5 -2.5 mas, the subsequent reprocessing of this data by van Leeuwen[12] maintained the individual data points, and delivered uncertainties in the range 0.8-1.0 mas (see Supplementary Table 1), improving the precision by a factor of ~4. The latter analysis provides for Beta Pictoris a parallax of 51.44±0.12 mas, corresponding to a distance of 19.44±0.05 pc, and a proper motion of 4.65±0.11 mas yr$^{-1}$ and 83.10±0.15 mas yr$^{-1}$ in the right ascension and declination directions (Figure 1; Table 1). The star is the largest member of an association of young stars, the Beta Pictoris moving group (BPMG), sharing the same origin and motion through space.

Beta Pictoris is included in the second data release[16] (DR2) of Gaia, and covers for this star 32 observations between October 1, 2014 and April 19, 2016, of which 30 were used for the astrometric solution. The individual measurements are not provided with DR2. For most stars observed by Hipparcos, Gaia is expected to eventually achieve a precision that is 100x better[17]. However, Beta Pictoris is such a bright star (V=3.86) that its images in the Gaia data are saturated, significantly degrading the achieved positional accuracy, resulting in DR2 uncertainties of 0.3 mas in each positional direction and parallax, and 0.7 mas yr$^{-1}$ in proper motion. However, since the Hipparcos and Gaia missions span a baseline of 24 years, the combined positional data provide a long-term proper motion measurement[16] with a precision of 0.02 mas yr$^{-1}$ (Table 1).

Figure 2 shows the heliocentric Hipparcos positional data after subtraction of the 24-yr Hipparcos-Gaia baseline. The residual proper motion is governed by the reflex motion of Beta Pictoris around the centre of mass of the system. The planet Beta Pictoris b is in a nearly edge-on orbit at a position angle of 212° (ref 19) consistent with the observed motion of the star in the Hipparcos data. The planet discovery was announced in 2009 from high-contrast images taken in 2003 (ref 2&3). Only after the planet reappeared on the other side (SW) of the star, it started to be intensely monitored, in particular with the new generation of high-contrast imaging instruments GPI on the Gemini Telescope and SPHERE on the Very Large Telescope[19-21]. It means that while about half of the orbit is well known, the orbital period is still poorly constrained to be between 20.2 to 26.3 years (ref 19). Since the reference epochs of Hipparcos (1991.25) and Gaia DR2 (2015.5) are separated by 24.25 years, the observations of the former mission fall in the well-determined part of the planet orbit, but the precise orbital phase is uncertain. Depending on the orbital period, the Hipparcos-Gaia baseline also contains a small part of the stellar reflex motion. E.g. for an orbital period of 22 years and planet mass of 10 M$_{Jup}$, the two reference epochs correspond to orbital phases that differ by 0.1, resulting in a planet-induced position difference of 1.0 mas and an effect on the Hipparcos-Gaia baseline of 1.0/24.25 = 0.04 mas yr$^{-1}$. This effect is taken into account in our calculations below.

Since the projected orientation of the orbit is well constrained[19], so is the expected direction of stellar reflex motion. Hence, we converted the positional data from Hipparcos to one-dimensional position measurements in this direction as function of time, which is shown in the right panel of Figure 2. The solid curves show the best-fit stellar motion for several trial orbital periods, indicating that short periods are not consistent with the Hipparcos data since this would result in significant acceleration in the stellar proper motion in the period 1990-1993, which is not observed. Figure 3 shows the constraints to the planet mass and orbital period. The green-shaded area indicates the 1σ uncertainty interval on the orbital period from the high-contrast imaging monitoring[19], and the contours the 1, 2, and 3σ limits on orbital period and mass from the study presented here. The planet mass is constrained to 11±2 $M_{Jup}$ (1σ), and the orbital period is likely to be >22.2 years (2σ limit). Previous upper limits to the planet mass of <20-30 $M_{Jup}$ from radial velocity monitoring[9,10] are consistent with the astrometric measurement presented here.

We compared our measurement with the mass estimates obtained from gas giant cooling models. The wide range of estimates in the literature are driven by different assumptions on the age of the planet, and the internal energy of the planet at time of formation. In particular, it is uncertain what fraction of the released gravitational energy of newly accreted material during formation is lost in the accretion shock on the planet surface or circumplanetary disk and not incorporated in the planetary structure, resulting in a range of so called "hot" start, "cold" start, and intermediate "warm" start models[5-8]. In addition, different age-determination methods give different stellar ages, and also it is not certain whether the planet was formed at exactly the same time as the star, or possibly a few millions of years later. Isochronal fits to low-mass BPMG members provide[22] an age of 8-20 Myr, while dynamical ages of the BPMG provide age estimates in the range 10-12 Myr (ref 23) to 20 Myr (ref 24). The lithium depletion boundary of the BPMG points[25] to an age of 21±4 Myr. The older the system and the lower the initial energy, the higher is the theoretically inferred mass of the planet[8]. This has led to mass-estimates for Beta Pictoris b of 4-11 $M_{Jup}$ (ref 26), 6-12 $M_{Jup}$ (ref. 3), 8-13 MJup (ref 27),10-17 Mjup (ref 25), but also recently 12.9±0.2 $M_{Jup}$ (ref 28). The most recent mass estimates are in line with the model-independent measurement presented here, implying that the underlying assumptions in the hot start models and the age estimates of the system are largely correct.

The ongoing data collection by Gaia will in the near future lead to a further improvement on the mass measurement of Beta Pictoris b by up to an order of magnitude, and the masses of other known young planets will be in reach[17,18]. This means that Gaia will form a great asset to further constrain early evolution scenarios. In addition, Gaia is poised[29] to find many new extrasolar planets.

**Acknowledgments**

This work was only possible because numerous scientists and engineers have devoted large parts of their careers to the design, construction, and successful operation of the Hipparcos and Gaia missions, and the analysis of their data. We are grateful. We thank Floor van Leeuwen also for discussions on his analysis of the Hipparcos data. This work has made use of data from the European Space Agency (ESA) mission Gaia (https://www.cosmos.esa.int/gaia), processed by the Gaia Data Processing and Analysis Consortium (DPAC, https://www.cosmos.esa.int/web/gaia/dpac/consortium). Funding for the DPAC has been provided by national institutions, in particular the institutions participating in the Gaia Multilateral Agreement. I.S. acknowledges funding from the European Research Council (ERC) under the European Union's Horizon 2020 research and innovation programme under grant agreement No 694513. A.B. acknowledges funding from the Netherlands Research School for Astronomy (NOVA).

**Authors Contributions**

I.S devised the general idea, conducted the main analysis, and wrote a first version of the manuscript. A.B contributed to the implementation and optimization of the analysis, including important details on the Gaia and Hipparcos data, and commented on the manuscript.

**Table 1**

The Hipparcos and Gaia DR2 astrometric data on Beta Pictoris (HIP 27321; Gaia DR2 4792774797545105664), showing the Right Ascension (RA), Declination (Dec), Parallax (Plx), and proper motion in RA and Declination direction (pmRA and pmDec).

| Hipparcos (ICRS; epoch 1991.25) | Gaia DR2 (ICRS; epoch 2015.5): |
|---|---|
| RA = $86.82118054°$ ± 0.10 mas | RA = $86.82123366°$ ± 0.314 mas |
| Dec = $-51.06671341°$ ± 0.11 mas | Dec = $-51.06614803°$ ± 0.342 mas |
| Plx = 51.44 ± 0.12 mas | Plx = 50.62 ± 0.33 mas |
| pmRA = 4.65 ± 0.11 mas | pmRA = 2.49 ± 0.68 mas |
| pmDec = 83.10 ± 0.15 mas | pmDec = 82.58 ± 0.68 mas |
| Hipparcos+Gaia (1991.25 – 2015.5) pmRA = 4.94 ± 0.02 mas pmDec = 83.93 ± 0.02 mas | |

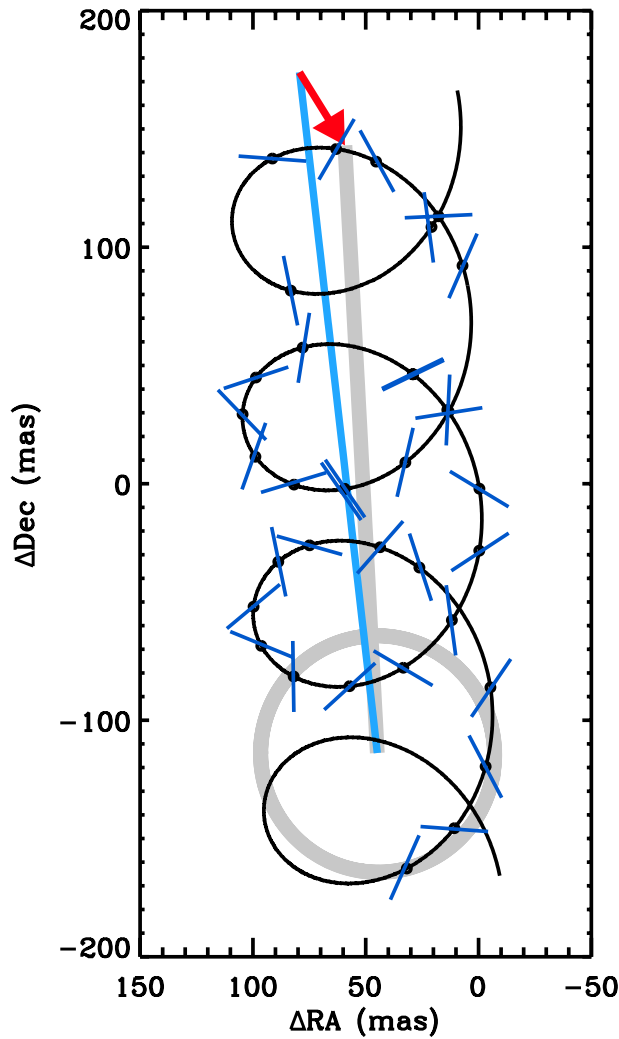

Fig. 1: The astrometric motion of the young star Beta Pictoris as shown by the Hipparcos data (1990-1993). The black curve is the best fit model to the parallax (caused by the Earth's yearly motion around the Sun) and the star's proper motion, with the points indicating the modelled positions for the one-dimensional Hipparcos positions (the dark blue lines are perpendicular to the scanning direction), which are averaged per satellite orbit for clarity and have typical error bars of <1 mas. The grey circle and line indicate the separate parallax and proper motion components. The light blue line shows the proper motion as determined from the 24-year Hipparcos-Gaia baseline, with the red arrow showing the difference with the internal Hipparcos proper motion caused by the reflex motion of the star around the system's barycentre. This difference is amplified by a factor 10 to make the effect visible.

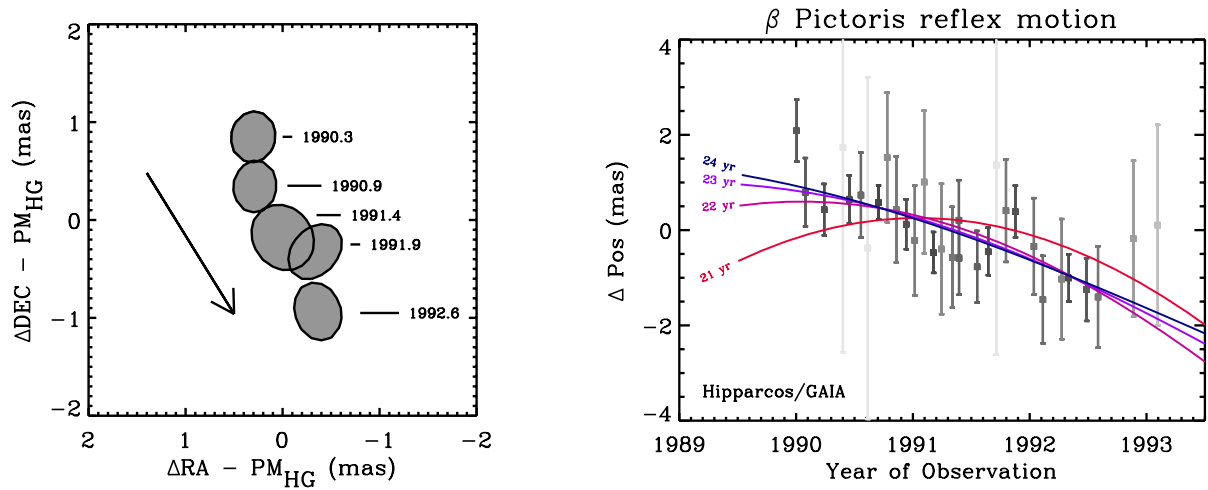

Fig 2. (panel A): The change in position of Beta Pictoris in the Hipparcos data relative to the 24-year Hipparcos-Gaia baseline. The one-dimensional positions taken during 35 Hipparcos spacecraft orbits were combined to 5 two-dimensional astrometric positions. The shaded areas indicate the 1σ uncertainty areas with their mean epochs. The arrow shows the expected direction for the movement of the star around the center of mass of the system. (panel B): The change in position of the star relative to the Hipparcos-Gaia baseline in the direction of the expected reflex motion direction of the star at 212° known from the orbital mapping of the planet. Positions averaged over a spacecraft orbit are shown which strongly vary in accuracy depending on the relative orientation of the one-dimensional astrometric measurement. Five measurements are omitted because their measurement vectors are near-perpendicular to the reflex motion movement. The colored lines show the best-fit velocity curves for orbital periods of 21 to 24 years. Planet orbital periods shorter than 22 years are excluded since they would show significantly more curvature than the Hipparcos data allows.

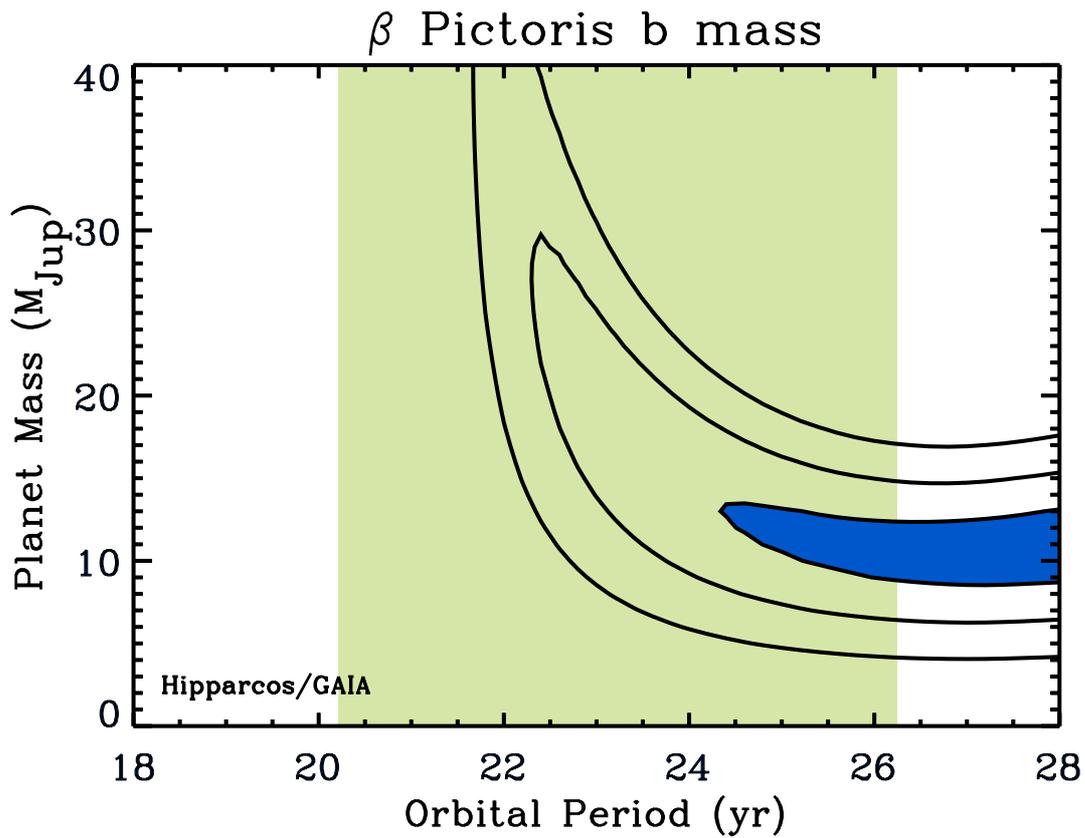

Fig. 3: Constraints on the mass and orbital period of the young exoplanet Beta Pictoris b. The contours show the single-parameter 1, 2, and 3σ uncertainty areas from the astrometric analysis presented here, with the blue area highlighting the 1σ region. Constraints on the orbital period from high-contrast imaging of the planet are shown in green. The planet is determined to have a mass of 11±2 $M_{Jup}$ (1σ), and the orbital period is likely to be >22.2 years (2σ limit). Longer orbital periods are permitted by the astrometric data, except for >>30 years when significant acceleration/deceleration in the motion of the star would be expected during the time of the Hipparcos observations.

**Methods: Data Analysis**

We collected the Hipparcos intermediate data products from the van Leeuwen 2007 reduction[14] from CD, which contain for each observation the spacecraft orbit and epoch, the cosine and sine of the scan orientation, the residual to the best fit solution, and the formal error (see Supplementary Table 1). All calculations were performed with custom-built IDL scripts. Figure 1 of the main text was produced using the `make_parallax_coords.pro` procedure[30], to convert the Hipparcos barycentric position of Beta Pictoris (epoch 1991.25), the intermediate data products, and the best-fit solution to the parallax and proper motion, in to a model of the on-sky geocentric movement of the star and the accompanying one-dimensional astrometric measurements. The Gaia (epoch=2015.5) barycentric position was extracted from the DR2 archive and combined with that of Hipparcos to determine the average proper motion of the star over the intermediate 24.25 years.

The 111 one-dimensional astrometric measurements of Hipparcos, taken during 35 spacecraft orbits, where combined into five two-dimensional positions (Figure 2, left panel), c.f. combining observations with seven different scanning directions per position. This was done by combining the individual measurements per spacecraft orbit to produce a [$\Delta$RA, $\Delta$Dec] grid of $\Delta\chi^2$ values assuming a Gaussian noise distribution along the scanning direction. These $\Delta\chi^2$ grid-maps were subsequently combined to $\Delta\chi^2=1$ contours for each of the five epochs, as shown in the left panel of Figure 2. It shows that the movement of the star is in the expected direction for the reflex motion induced by planet Beta Pictoris b, as indicated by the arrow.

Since the direction of the reflex motion of the host star is well determined[19] ($\varphi = 212°$), we converted each orbit-averaged observation into a position-measurement along this direction and adjusted its associated uncertainty, which depends on 1/cosine of the angle relative to the scanning direction. This is shown in the right panel of Figure 2, where the data points have grey scales depending on their uncertainty to emphasis the data with the smallest errors. Five measurements were omitted because their scanning directions were near-perpendicular leading to very large errors >10 mas. The shape of the orbit of the planet is well determined for the Hipparcos 1990-1993 observing interval, because it is approximately one planet orbit (20 – 26 yrs) before the current era (2010 – present) during which the planet position has been frequently monitored (see Supplementary Figure 1). Therefore, although the orbital period is not well constrained, the best-fit[19] solution to its orbit can be used to fit the reflex motion of the star, modulo a scaling depending on the planet/star mass ratio and a phase shift governed by the assumed orbital period. This is shown for trial orbital periods between 21 and 24 years in the right panel of Figure 2. For a short orbital period of 21 years, the Hipparcos observations fall around the epoch of longest elongation, implying a strong acceleration/deceleration of its angular motion – which is excluded by the Hipparcos data.

We subsequently performed a chi-squared analysis to the data presented in the right panel of Figure 2 over a fine grid of orbital period and planet mass (assuming a stellar mass[31] of M=1.75 $M_{Sun}$), of which the results are shown in Figure 3, with the 1, 2, and 3σ uncertainty intervals indicated by the contours. The green area is the 1σ uncertainty interval for the orbital period as determined from high-contrast imaging of the planet[19].

We checked to what extent the deviation in the Gaia parallax from that of Hipparcos of -0.82 mas can influence the Hipparcos-Gaia proper motion measurement. Gaia-DR2 quote a negligible RA-parallax correlation, and a correlation in Dec-parallax of +0.248, implying a deviation in position of -0.82 x 0.248 = -0.2 mas. This results in a change in the Hipparcos-Gaia proper motion, which has a baseline of 24.25 years, of only -0.2/24.25= -0.008 mas $yr^{-1}$.

**Data availability.** The data that support the plots within this paper and other findings are available in Table 1 of the Supplementary Information, and from the Gaia data archive https://gea.esac.esa.int/archive/

**Code availability:** All analyses were performed with custom-built IDL scripts.

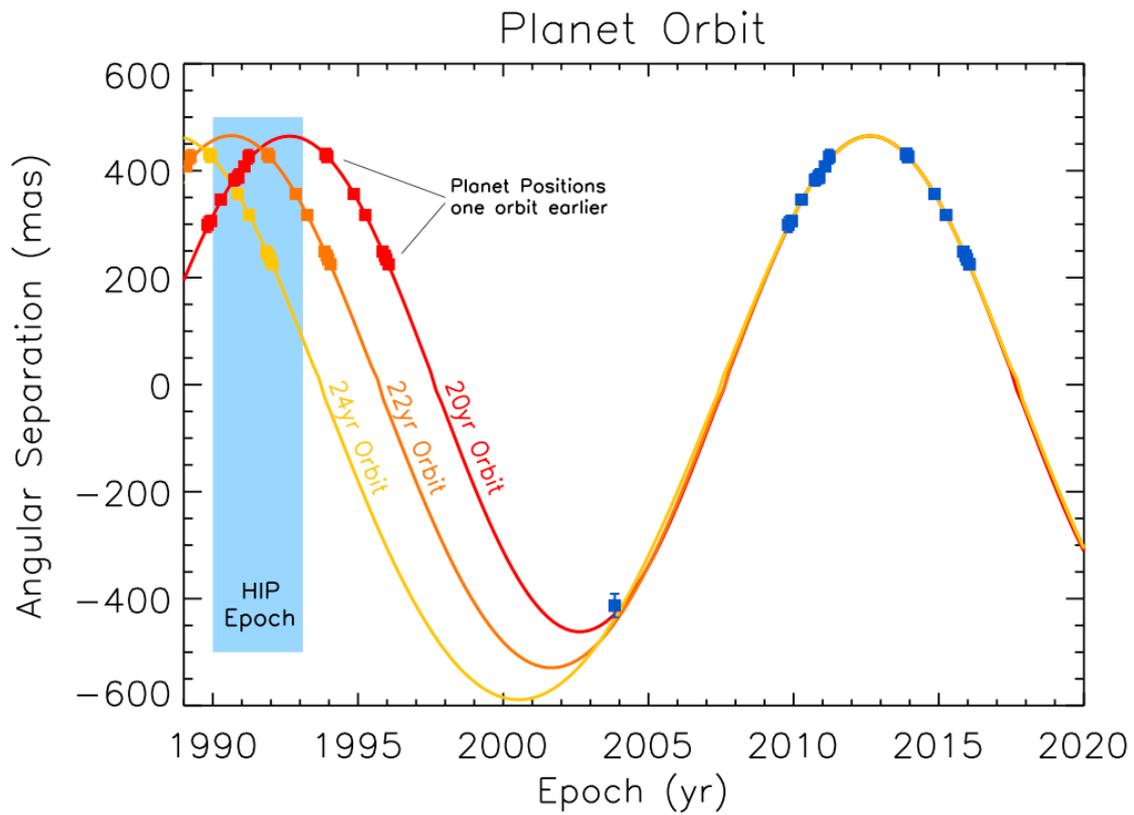

Supplementary Figure 1: The angular separation of the planet Beta Pictoris b for different orbital solutions with periods of 20 yrs (red), 22 yrs (orange), and 24 yrs (yellow). The blue squares are the observed positions as collected by ref 17. Those in red, orange, and yellow are the same data points, but moved by one orbit. It shows that except for a shift in time depending on the assumed orbital period, the planet positions are well determined for the Hipparcos epochs (light-blue bar; 1990 – 1993), since observations constrain this part of the orbit very well.

| Supplementary Table 1 Hipparcos intermediate data products from the *van Leeuwen* re-analysis of Beta Pictoris (HIP 27321). | | | | | | |
|---|---|---|---|---|---|---|
| Orbit Number | Epoch (yr -1991.25) | Parallax Factor | Cosine of scan orientation | Sine of scan orientation | Residual to best-fit solution (mas) | Formal error (mas) |
| 133 | -1.245 | 0.624 | -0.9065 | -0.4222 | -1.00 | 0.81 |
| 133 | -1.245 | 0.626 | -0.9050 | -0.4254 | -0.31 | 0.80 |
| 133 | -1.245 | 0.622 | -0.9076 | -0.4198 | -1.66 | 0.78 |
| 194 | -1.170 | -0.651 | -0.0680 | 0.9977 | 0.39 | 0.78 |
| 194 | -1.170 | -0.649 | -0.0716 | 0.9974 | -0.79 | 0.86 |
| 194 | -1.170 | -0.648 | -0.0723 | 0.9974 | -0.64 | 0.87 |
| 257 | -1.094 | 0.671 | -0.8760 | 0.4823 | 0.01 | 0.80 |
| 257 | -1.094 | 0.674 | -0.8778 | 0.4790 | -0.23 | 0.78 |
| 257 | -1.094 | 0.671 | -0.8756 | 0.4830 | -0.20 | 0.85 |
| 327 | -1.009 | -0.679 | 0.8136 | 0.5814 | 0.19 | 0.75 |
| 327 | -1.009 | -0.680 | 0.8147 | 0.5799 | 0.02 | 0.82 |
| 327 | -1.009 | -0.679 | 0.8137 | 0.5813 | -1.26 | 0.81 |
| 458 | -0.849 | -0.641 | 0.9457 | -0.3250 | 0.51 | 0.91 |
| 458 | -0.849 | -0.641 | 0.9457 | -0.3250 | -1.01 | 0.82 |
| 458 | -0.849 | -0.640 | 0.9461 | -0.3240 | 0.25 | 1.06 |
| 458 | -0.849 | -0.641 | 0.9457 | -0.3250 | -1.17 | 0.84 |
| 458 | -0.849 | -0.640 | 0.9458 | -0.3247 | -1.04 | 0.80 |
| 503 | -0.794 | 0.637 | 0.7353 | 0.6777 | 0.75 | 0.95 |
| 503 | -0.794 | 0.636 | 0.7367 | 0.6762 | -0.01 | 0.82 |
| 503 | -0.794 | 0.636 | 0.7365 | 0.6764 | -0.51 | 1.10 |
| 503 | -0.794 | 0.636 | 0.7365 | 0.6765 | -0.34 | 0.96 |
| 585 | -0.695 | -0.656 | 0.2592 | -0.9658 | -0.22 | 0.70 |
| 585 | -0.695 | -0.655 | 0.2602 | -0.9656 | -0.52 | 0.86 |
| 585 | -0.695 | -0.655 | 0.2609 | -0.9654 | 0.73 | 0.81 |
| 634 | -0.636 | 0.674 | 0.9777 | -0.2102 | -0.19 | 0.80 |
| 709 | -0.544 | -0.687 | -0.6238 | -0.7816 | 0.02 | 0.67 |
| 709 | -0.544 | -0.685 | -0.6225 | -0.7826 | 0.09 | 0.73 |
| 709 | -0.544 | -0.685 | -0.6216 | -0.7833 | -0.41 | 0.72 |
| 709 | -0.544 | -0.686 | -0.6226 | -0.7826 | -0.21 | 0.76 |
| 770 | -0.470 | 0.673 | 0.3654 | -0.9308 | -1.15 | 0.84 |
| 770 | -0.470 | 0.672 | 0.3640 | -0.9314 | -0.04 | 0.84 |
| 834 | -0.392 | -0.640 | -0.9999 | 0.0141 | 0.66 | 0.78 |
| 834 | -0.392 | -0.643 | -0.9998 | 0.0182 | -0.21 | 0.81 |
| 834 | -0.392 | -0.640 | -0.9999 | 0.0145 | -1.21 | 0.84 |
| 834 | -0.392 | -0.642 | -0.9998 | 0.0177 | 0.20 | 0.76 |
| 904 | -0.307 | 0.617 | -0.6578 | -0.7532 | 0.20 | 0.71 |
| 904 | -0.307 | 0.617 | -0.6584 | -0.7526 | 0.05 | 0.74 |
| 964 | -0.234 | -0.625 | -0.4897 | 0.8719 | -0.86 | 0.78 |
| 964 | -0.234 | -0.623 | -0.4919 | 0.8706 | 0.36 | 1.24 |
| 964 | -0.234 | -0.623 | -0.4930 | 0.8700 | -1.03 | 0.86 |
| 964 | -0.234 | -0.628 | -0.4859 | 0.8740 | -0.30 | 0.86 |
| 964 | -0.234 | -0.625 | -0.4904 | 0.8715 | 0.82 | 0.83 |
| 1031 | -0.153 | 0.655 | -0.9905 | 0.1376 | 0.10 | 0.73 |
| 1031 | -0.153 | 0.651 | -0.9899 | 0.1420 | -0.37 | 0.88 |
| 1031 | -0.153 | 0.654 | -0.9903 | 0.1389 | -0.49 | 0.84 |
| 1031 | -0.153 | 0.652 | -0.9901 | 0.1404 | -0.85 | 0.84 |
| 1097 | -0.073 | -0.675 | 0.5283 | 0.8491 | -2.09 | 1.11 |
| 1097 | -0.073 | -0.679 | 0.5330 | 0.8461 | -0.90 | 1.08 |
| 1097 | -0.073 | -0.677 | 0.5301 | 0.8479 | -0.25 | 0.89 |
| 1097 | -0.073 | -0.677 | 0.5305 | 0.8477 | 0.14 | 1.16 |
| 1097 | -0.073 | -0.678 | 0.5318 | 0.8469 | -0.05 | 0.80 |
| 1153 | -0.005 | 0.675 | -0.4994 | 0.8664 | 0.04 | 0.83 |
| 1153 | -0.005 | 0.676 | -0.5006 | 0.8657 | 0.21 | 0.91 |
| 1153 | -0.005 | 0.675 | -0.4995 | 0.8663 | 0.73 | 1.15 |
| 1153 | -0.005 | 0.675 | -0.4995 | 0.8663 | -1.65 | 0.98 |
| 1229 | 0.088 | -0.653 | 0.9986 | 0.0528 | 0.30 | 0.86 |
| 1229 | 0.088 | -0.654 | 0.9987 | 0.0518 | -0.53 | 1.04 |
| 1229 | 0.088 | -0.654 | 0.9987 | 0.0513 | -0.11 | 1.00 |
| 1229 | 0.088 | -0.654 | 0.9987 | 0.0516 | -0.84 | 0.83 |
| 1275 | 0.144 | 0.636 | 0.4213 | 0.9069 | 0.34 | 0.85 |
| 1276 | 0.145 | 0.644 | 0.4193 | 0.9078 | -0.45 | 0.77 |
| 1404 | 0.300 | 0.657 | 0.9862 | 0.1654 | -0.43 | 1.06 |
| 1404 | 0.300 | 0.656 | 0.9864 | 0.1643 | 1.55 | 0.96 |
| 1404 | 0.300 | 0.657 | 0.9861 | 0.1661 | 0.71 | 0.90 |
| 1404 | 0.300 | 0.657 | 0.9861 | 0.1659 | -0.75 | 0.79 |
| 1404 | 0.300 | 0.658 | 0.9860 | 0.1665 | -2.69 | 1.19 |
| 1404 | 0.300 | 0.658 | 0.9861 | 0.1664 | -1.40 | 1.09 |
| 1482 | 0.395 | -0.683 | -0.3021 | -0.9533 | 0.38 | 0.85 |
| 1482 | 0.395 | -0.683 | -0.3023 | -0.9532 | -1.30 | 1.06 |
| 1482 | 0.395 | -0.682 | -0.3009 | -0.9536 | 0.11 | 0.83 |
| 1482 | 0.395 | -0.683 | -0.3023 | -0.9532 | 1.19 | 1.15 |
| 1539 | 0.464 | 0.690 | 0.7038 | -0.7104 | -0.50 | 0.85 |
| 1539 | 0.464 | 0.691 | 0.7044 | -0.7098 | 1.08 | 1.22 |
| 1539 | 0.464 | 0.689 | 0.7033 | -0.7109 | -0.49 | 0.88 |
| 1539 | 0.464 | 0.689 | 0.7027 | -0.7115 | 1.45 | 1.07 |
| 1607 | 0.547 | -0.664 | -0.9372 | -0.3487 | -1.27 | 1.23 |
| 1607 | 0.547 | -0.665 | -0.9374 | -0.3483 | -0.21 | 0.93 |
| 1674 | 0.628 | 0.631 | -0.2787 | -0.9604 | -1.03 | 1.07 |
| 1674 | 0.628 | 0.634 | -0.2749 | -0.9615 | -1.85 | 1.17 |
| 1674 | 0.628 | 0.633 | -0.2756 | -0.9613 | 0.30 | 1.21 |
| 1674 | 0.628 | 0.631 | -0.2785 | -0.9604 | -0.98 | 0.83 |
| 1734 | 0.701 | -0.623 | -0.8109 | 0.5853 | 1.93 | 0.77 |
| 1734 | 0.701 | -0.620 | -0.8130 | 0.5823 | -0.85 | 1.18 |
| 1734 | 0.701 | -0.622 | -0.8121 | 0.5835 | -0.05 | 1.19 |
| 1735 | 0.702 | -0.621 | -0.8083 | 0.5888 | -1.82 | 1.18 |
| 1804 | 0.786 | 0.630 | -0.9717 | -0.2363 | 0.09 | 0.77 |
| 1804 | 0.786 | 0.630 | -0.9715 | -0.2369 | -0.34 | 0.77 |
| 1867 | 0.863 | -0.664 | 0.1487 | 0.9889 | -0.58 | 0.76 |
| 2000 | 1.024 | -0.669 | 0.9088 | 0.4172 | -0.23 | 0.91 |
| 2049 | 1.084 | 0.650 | 0.0446 | 0.9990 | 0.04 | 0.73 |
| 2049 | 1.084 | 0.650 | 0.0441 | 0.9990 | -0.02 | 0.77 |
| 2049 | 1.084 | 0.650 | 0.0440 | 0.9990 | -0.24 | 0.72 |
| 2049 | 1.084 | 0.650 | 0.0441 | 0.9990 | 0.38 | 0.72 |
| 2130 | 1.182 | -0.633 | 0.8661 | -0.4999 | -0.27 | 0.70 |
| 2130 | 1.182 | -0.633 | 0.8659 | -0.5002 | -0.17 | 0.91 |
| 2130 | 1.182 | -0.632 | 0.8661 | -0.4998 | -0.16 | 0.90 |
| 2130 | 1.182 | -0.632 | 0.8661 | -0.4999 | -1.14 | 0.98 |
| 2175 | 1.237 | 0.640 | 0.8524 | 0.5229 | -0.59 | 1.30 |
| 2175 | 1.237 | 0.640 | 0.8523 | 0.5231 | -1.31 | 0.97 |
| 2175 | 1.237 | 0.638 | 0.8536 | 0.5210 | 0.17 | 1.05 |
| 2175 | 1.237 | 0.638 | 0.8537 | 0.5207 | 0.88 | 1.18 |
| 2256 | 1.335 | -0.663 | 0.0709 | -0.9975 | -0.22 | 1.26 |
| 2256 | 1.335 | -0.664 | 0.0702 | -0.9975 | 0.26 | 0.97 |
| 2506 | 1.639 | -0.632 | -0.9775 | 0.2107 | -0.51 | 0.76 |
| 2506 | 1.639 | -0.628 | -0.9787 | 0.2055 | -0.17 | 0.92 |
| 2506 | 1.639 | -0.630 | -0.9781 | 0.2081 | 0.04 | 0.84 |
| 2506 | 1.639 | -0.633 | -0.9771 | 0.2128 | 0.39 | 0.89 |
| 2506 | 1.639 | -0.628 | -0.9786 | 0.2058 | -0.28 | 0.81 |
| 2677 | 1.846 | 0.652 | -0.9908 | 0.1352 | -0.03 | 0.85 |
| 2677 | 1.846 | 0.651 | -0.9904 | 0.1380 | -2.97 | 1.63 |
| 2677 | 1.846 | 0.651 | -0.9906 | 0.1364 | -0.69 | 0.83 |